\documentclass[12pt,preprint]{aastex}

\newcommand{\kev}{keV}
\newcommand{\etal}{et al.}

\newcommand{\microjy}{$\mu$Jy}
\newcommand{\sfr}{M$_{\odot}$~yr$^{-1}$}
\newcommand{\lx}{$L_{X}$}

\shorttitle{AGNs, Star Formation and Deep Radio Surveys}
\shortauthors{Ballantyne}

\begin{document}

\title{The Contribution of Active Galactic Nuclei to the Microjansky
  Radio Population}


\author{D. R. Ballantyne}
\affil{Center for Relativistic Astrophysics, School of Physics,
  Georgia Institute of Technology, Atlanta, GA 30332;
  david.ballantyne@physics.gatech.edu}

\begin{abstract}
A X-ray background synthesis model is used to calculate the
contribution of Active Galactic Nuclei (AGNs) to the 1.4~GHz number
counts between 100~nJy and 10~mJy. The number counts are broken down
into contributions from radio-quiet and radio-loud AGNs, obscured and
unobscured AGNs, and for different ranges in redshift and 2--10~\kev\
X-ray luminosity, $L_{X}$. Compton-thick AGNs are included, but only
to the level required to fit the peak of the X-ray background. The
predicted radio counts show that the \microjy\ AGN population will be
dominated by obscured, radio-quiet Seyfert galaxies with $\log L_{X} <
43$, and spanning $0 < z \la 3$. However, depending on the exact
relationship between the radio and X-ray luminosities in radio-quiet AGNs,
additional radio flux due to star-formation within AGN host
galaxies may be necessary in order
to match the observed AGN counts at a flux density of $\sim 50$~\microjy. The
star-formation rates (SFR) required are modest, only $\sim 3$~\sfr,
assuming a constant rate with $z$ and $L_X$. A more observationally
and theoretically motivated relationship, where the SFR~$\propto
(1+z)^{1.76}(\log L_{X}-40)^{3.5}$, will also account for the observed
counts. The \microjy\ AGN population will provide a very clean sample
to trace the accretion and galactic star-formation histories of
Seyfert galaxies over a significant fraction of cosmic time.
\end{abstract}


\section{Introduction}
\label{sect:intro}
The identification of powerful radio sources with quasi-stellar
objects in the 1960s \citep{ms63,hms63,sch63} ultimately led to the
realization that Active Galactic Nuclei (AGNs) were the locations of
accreting supermassive black holes \citep[e.g.,][]{salp64,lb69}. Since
that time, radio observations have continued to be an important probe
of the AGN phenomenon, especially in the investigation of relativistic
jets emanating from the central engine
\citep[e.g.,][]{bp84,ko88,odea02}. The next decade will see a major
leap forward of the capabilities of centimeter wave radio
instrumentation with the development of the Expanded-VLA (EVLA;
\citealt{nap06}) and, ultimately, the Square Kilometer
Array\footnote{\texttt{http://www.skatelescope.org}}. The increase in
sensitivity afforded by these new instruments will probe the radio
population down to flux densities less than a \microjy, opening a new
window into the study of AGN evolution and the growth of supermassive
black holes.

At flux densities $S \ga 1$~mJy, the radio source population is
dominated by radio-loud AGNs, and it has long been known that the
number count distribution indicates strong cosmic evolution
\citep[e.g.,][]{long66,cond84,cond89}. However, the radio number
counts significantly flatten at fluxes below 1~mJy
\citep[e.g.,][]{cond84,wind85,hopk98,rich00,smg04,simp06,kell08}
indicating either a sudden change in evolutionary properties, or, more
likely, the appearance of a new population of sources such as emission
from star-forming galaxies
\citep[e.g.,][]{wind95,benn93,rich00,smg04}. Strong radio emission
from star-forming regions is expected as supernova remnants will
accelerate cosmic rays which will radiate synchrotron emission in
local magnetic fields. At low redshifts, the radio and far-infrared
powers of star-forming galaxies are tightly correlated, allowing for
the radio luminosity to act as a star-formation rate (SFR) indicator
\citep{cond92,yrc01,bell03}. This correlation seems to hold at higher
redshifts \citep{bes08, ibar08, ga09}.

In recent years, several surveys have continued to probe the radio
source population to ever-smaller values
\citep[e.g.,][]{hopk03,simp06,sch07,kell08}, with the deepest VLA
surveys reaching rms noise levels of only a few \microjy\ per
beam. Matching the radio sources to complementary optical, X-ray and
infrared surveys has allowed the properties of many of the faintest
radio sources to be estimated \citep{sey08,smo08,pad09}. The results
of these studies, while still with large error-bars, indicate that
radio-quiet AGNs may make up to one-third of the radio source
population below 100~\microjy\ \citep{pad09}. As $\sim$90\% of AGNs
are radio-quiet, these results imply that the deep radio surveys are
now uncovering the dominant population of active galaxies.

The last decade has provided a wealth of new data tracing the
evolution of supermassive black holes and their impact on galaxy
formation
\citep[e.g.,][]{kauff03,grog05,pierce07,silv08,gab09,sch09,hick09},
but it has proved difficult to simultaneously study the AGN and host
galaxy. Galaxies are much less luminous than AGNs at X-ray energies,
so deep X-ray observations are very efficient at finding all but the
most heavily obscured AGNs over a wide range of redshift and
luminosity \citep{mush04,bh05}. These surveys have showed that the
vast majority of AGNs in the Universe are absorbed, and that AGNs
evolve in the same form of `cosmic downsizing' as the galaxy
population \citep[e.g.,][]{ueda03, bar05}. However, aside from
indirect inferences using the evolution of the X-ray absorber
\citep{bem06}, it is very challenging to glean much information about
the affect of the AGN on the host galaxy through X-ray
observations. Optical, ultraviolet, and infrared wavelengths allow
study of both the host galaxy and AGN, but the extinction and
re-emission properties of dust in the host galaxy severely complicate
the interpretation when, as is nearly always the case, the galaxy is
not spatially resolved \citep[e.g.,][]{mfc02,don08}. The radio
regime is unique in that it provides a nearly uncorrupted view of both
processes. All AGNs produce some form of non-thermal nuclear radio
emission which, although weak, can be identified by its high
brightness temperature \citep[e.g.,][]{hn92,blu96,kuk98} and is immune
to attenuation by any circumnuclear torus. Radio-emission from the
host galaxy will provide a similarly unabsorbed view into ongoing
star-formation. Thus, combining studies of the \microjy\ radio-quiet
AGN population with other multiwavelength surveys will allow a
relatively clean look into the combined problem of AGN and galaxy
evolution.

In this paper, we make use of a X-ray background synthesis model to
predict the number counts of radio-quiet AGNs down to a flux density
of 100 nJy. The calculations break down the contribution of AGNs from
different redshifts and nuclear luminosities. In addition, we add in
radio-emission from various levels of star-formation in the AGN host
galaxies to explore the impact on the predicted number counts. The
next section describes the calculations necessary to move from a X-ray
background synthesis model to predicting the radio number
counts. Section~\ref{sect:res} then presents and describes the
results, including the effects of adding in star-formation. The
results are discussed and summarized in Section~4. The following
$\Lambda$-dominated cosmology is assumed in this paper:
$H_0=70$~km~s$^{-1}$~Mpc$^{-1}$, $\Omega_{\Lambda}=0.7$, and
$\Omega_{m}=0.3$ \citep{spe03}. The flux density of a radio source has
a spectral index $\alpha$, so that $S \propto \nu^{\alpha}$.

\section{Calculations}
\label{sect:calc}
Previously, \citet{jr04} and \citet{wil08} estimated the radio-quiet
AGN contribution to the radio counts by employing the \citet{ueda03}
hard X-ray luminosity function, multiplying by a constant to correct
for Compton-thick AGNs, and then employing a \textit{ROSAT}-based
correlation \citep{brink00} to convert from X-ray to radio
luminosity. In both these papers, they use separate radio-derived
luminosity functions to account for radio-loud AGNs. Below, we present
a new method to estimate the radio-quiet AGN number counts that will
improve on this previous work in several ways. First, we base our
calculations on a fit to the X-ray background. This allows the peak
spectral intensity of the background at $\sim 30$~\kev\ to be used as
a constraint on the missing Compton-thick fraction, and has the added
feature of allowing us to break down the number counts into
contributions from different groups such as obscured AGNs, or ones in
a specific redshift or luminosity range. Furthermore, we account for a
radio-loud fraction of AGN. The X-ray luminosity function counts both
radio-loud and radio-quiet AGNs, so the radio-loud fraction must be
subtracted from the total before predicting the radio-quiet
counts. Finally, we convert from X-ray to radio luminosity
using a luminosity-dependent conversion factor derived from
high-resolution radio and hard X-ray observations of the nuclear
regions of nearby AGNs. While clearly there remain several
uncertainties in the steps described below, the changes outlined
above should result in a more accurate prediction of the AGN
contribution to the deep radio counts. In these calculations, an AGN
is assumed to be obscured if it is attenuated by a hydrogen column
$N_{\mathrm{H}} \geq 10^{22}$~cm$^{-2}$.

\subsection{The X-ray Background Synthesis Model}
\label{sub:xrb}
The hard X-ray background encodes within it the entire history of
accretion onto supermassive black holes, independent of whether or not
an individual AGN is radio-loud or radio-quiet.
A detailed description of the X-ray background synthesis model
employed here is beyond the scope of the current paper, and, indeed, many
aspects of the model such as the X-ray spectral shapes, and the
distribution of column densities have no impact on the predicted radio
number counts. For this application the important ingredients of the
model are the absorption-corrected hard X-ray luminosity function, and
the obscured-to-unobscured AGN ratio (i.e., the ratio of type 2 AGNs to
type 1 AGNs). The \citet{ueda03} hard X-ray luminosity function is the
basis for the calculations presented here.

\begin{figure}
\epsscale{1.0}
\plotone{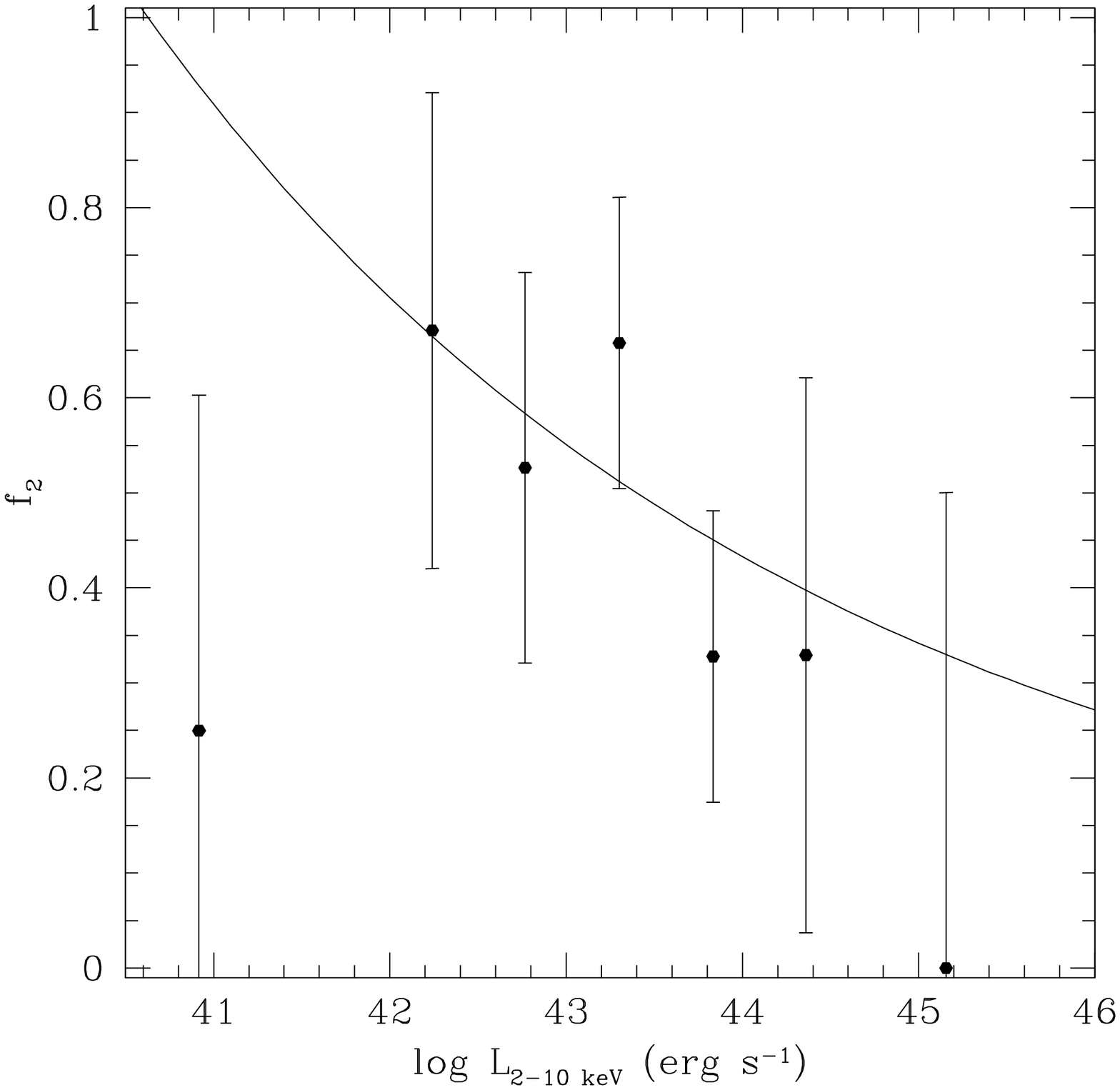}
\caption{The fraction of type 2 AGNs as a function of \lx\ as
  determined by the \textit{Swift}/BAT survey of \citet{tuel08}. The observed
  correlation between \lx\ and $L_{\mathrm{14-195\ keV}}$
  \citep{wint09} was used to convert the luminosities to the \lx\ band. The
  solid line is a power-law fit to the data assuming a 4:1
  type 2/type 1 ratio at $\log L_{X} = 41.5$. The best
  fit has a reduced $\chi^2=0.97$ and follows $f_2 \propto \log
  L_{X}^{-10.5}$.}
\label{fig:batf2}
\end{figure}

The AGN type 2/type 1 ratio is observed to significantly decrease with
AGN luminosity \citep{ueda03,laf05,simp05,ak06,tkd08}.  The explanation for this effect is not entirely
understood, but is most likely related to the increased outward
radiation pressure produced at high AGN luminosities \citep{law91,simp05}. The best
determination of the local ($z \sim 0$) dependence of the AGN type
2/type 1 ratio with luminosity was produced by the \textit{Swift}/BAT
detector in the 14--195~\kev\ band \citep{tuel08}. AGN selection at these energies is
expected to be completely unbiased to objects with Compton-thin
absorbers. Figure~\ref{fig:batf2} plots the BAT AGN type~2 fractions, $f_2$
as a function of \lx$=L_{\mathrm{2-10\ keV}}$ where the conversion $\log L_{X} \approx 1.06 \log L_{\mathrm{14-195\ keV}}-3.08$ found by
\citet{wint09} was used to convert the BAT luminosities to \lx. 
The plot shows a strong trend of an increasing fraction of obscured
AGNs as the nuclear luminosity decreases. The lowest luminosity point
indicates a much lower obscured fraction than expected by this trend,
but lies at such low luminosity that it is likely contaminated by
LINERs and other low-luminosity AGNs that have significantly different
nuclear environments and evolutionary histories than rapidly
accreting AGNs at higher luminosities \citep{ho08}. In practice, the hard
X-ray luminosity functions are integrated down to $\log
L_{X} = 41.5$. A power-law of the form $f_2 \propto
L_{X}^{-\beta}$ was fit to the data with the
normalization set by the type 2/type 1 ratio at $\log
L_{X} = 41.5$. The lowest luminosity datum was
included in the fit, but its removal makes a negligible impact on the
results. Assuming an obscured-to-unobscured ratio of 4:1 at $\log
L_{X} = 41.5$, the best-fit (reduced $\chi^2=0.97$)
was found for $\beta=10.5$. The predicted radio number counts decrease
by only 6\% if the smooth step function shown by \citet{tuel08} is
used to describe the variation in $f_2$.

This description of the AGN type 2 fraction is based on
observations of local AGNs, but there is now accumulating evidence that
$f_2$ also increases with redshift as $(1+z)^{\gamma}$
\citep{laf05,bem06,tu06,has08}. This finding is still controversial \citep{dp06}, and the data is not
yet at the point to determine if there is a luminosity dependence to
$\gamma$. Here, we assume $\gamma=0.4$, a value consistent with the
various estimates, and that the AGN type 2 fraction evolves out to
$z=1$, where, in analogy with the cosmic star-formation rate density
\citep[e.g.,][]{hopk04,hb06}, it flattens to $\gamma=0$. If we instead
choose $\gamma=0.62$ with the evolution continuing to $z=2$
\citep{has08}, then the predicted radio counts increase by only 8\%.

The hard X-ray luminosity functions derived in the literature are
based on observations at energies less than $\sim 10$~\kev, and so are
insensitive to Compton-thick AGNs, those with absorbing
column densities $N_{\mathrm{H}} \ga 10^{24}$~cm$^{-2}$. Therefore, the unknown
fraction of Compton-thick AGNs has to be included `by hand' into any
X-ray background model in order to fit the peak of the spectrum at
$\sim 30$~\kev. For the \citet{ueda03} X-ray luminosity
function, it was found that a Compton-thick fraction equal to the
fraction of Compton-thin type 2 AGNs provided a very good fit to the
peak of the X-ray background. That is, at every $z$ and \lx, the
Compton-thick AGN fraction was equal to $f_2$. The implicit assumption
is that the Compton-thick population evolves in the same manner as the
Compton-thin AGNs.

\subsection{Translating the X-rays to Radio}
\label{sub:radio}
The X-ray background model provides the number density of obscured and
unobscured AGNs as a function of $z$ and \lx. To calculate how these
AGNs appear in the radio sky, the X-ray luminosity must first be
translated into radio luminosities, which, for radio-quiet AGNs, is
not necessarily straightforward. Previous work by \citet{jr04} and
\citet{wil08} made use of the correlation published by \citet{brink00}
based on AGNs selected through cross-correlating the \textit{ROSAT}
All-Sky Survey and the FIRST 1.4~GHz radio catalogue. However, it is not
clear if this correlation is valid over a wide range of AGN
luminosities. The \textit{ROSAT} All-Sky Survey was a shallow, soft
X-ray survey and the sources selected by \citet{brink00} are typically
unobscured, luminous quasars. These objects lie at substantial
redshifts ($z > 0.1$) and therefore its possible that the radio
emission in the FIRST beam may be contaminated by star-formation in
the host galaxies. As we are interested in the relationship between
the core X-ray and radio luminosities, it would be best to use results
based on observations of lower-luminosity radio-quiet AGNs that have
isolated the nuclear radio emission.

Therefore, we consider the results of \citet{tw03} and \citet{pan07} who studied the X-ray and
radio nuclear luminosities of nearby AGNs with high resolution
observations. The \citet{pan07} sample was limited to low luminosity
Seyfert galaxies and LINERs with $\log L_{X} < 43$, with many of the
AGNs accreting at very low rates with respect to the Eddington
limit. This is problematic because it is now known that radio-loudness
is more common at low accretion rates \citep{tw03,pan07,sik07}, so a
sample of low luminosity AGN may be biased toward high radio
luminosities. In contrast, \citet{tw03} plot the AGN radio to X-ray
luminosity ratio as a function of $L_X$ over a wide range of X-ray
luminosities.  Although there is a scatter of $\sim 0.5$ dex, they
find that it decreases with $L_X$ until it reaches a minimum at $\log
L_X \approx 43$--$44$ before increasing again. Defining $R_X = \log
\nu L_{\nu}\mathrm{(5\ GHz)}/\log L_{X}$ we quantify this behavior as
\begin{equation}
\label{eq:tw}
R_X = \left \{ \begin{array}{ll}
            -0.67\log L_X + 23.67 & 41.5 \leq \log L_X \leq 43\\
             -5 & 43 < \log L_X \leq 44 \\
            \log L_X -49 & 44 < \log L_X \leq 45 \\
             -4 & \log L_X > 45
            \end {array}
      \right.
\end{equation}
This is the relationship used to convert from the X-ray to the rest-frame
5~GHz radio luminosity. To calculate the observed flux at 1.4~GHz, a spectral index of $\alpha=-0.7$ is
assumed \citep{kuk98}. In Sect.~\ref{sub:panessa}, we will show how the
results change if the \citet{pan07} relationship is used for all
luminosities.

The X-ray to radio conversion for the radio-loud population is also
derived from \citet{tw03}: $\nu L_{\nu}\mathrm{(5\ GHz)} \approx 10^{-2}
L_{X}$. This proportionality factor has a scatter of
$\sim 1$~dex, but, as is seen below, the radio-loud population does
not impact our conclusions, as they comprise a negligible component
of the radio population at \microjy\ levels. A spectral index of
$\alpha=-0.8$ is assumed to calculate the observed 1.4~GHz fluxes of
radio-loud AGNs.

The small fraction of radio-loud AGNs has to be removed from the entire
X-ray selected population to concentrate on the properties of the
dominant radio-quiet AGNs below 100~\microjy. As mentioned above, radio-loud AGNs are most
common at high luminosity (the radio-loud quasars) and at very low
nuclear luminosities, but these latter objects are not important
to the X-ray background and are omitted here. At every $z$, the radio-loud fraction
is assumed to rise exponentially from $\log L_{X}=41.5$ to $46$: $f_{\mathrm{RL}}
= c \exp(\log L_{X} - 40) + b$. A decent fit to the radio number
counts between 1 and 10~mJy was found assuming
$f_{\mathrm{RL}}=0.0175$ at $\log L_{X}=41.5$ and
$f_{\mathrm{RL}}=0.10$ at $\log L_{X} \geq 46$. These values are
consistent with the canonical AGN radio-loud fraction of 10--20\% \citep[e.g.,][]{dbk09}.

The 1.4~GHz radio number counts are commonly displayed as the
differential counts, $dN/dS$. This can be related back to the X-ray
band by
\begin{equation}
\label{eq:dnds}
{dN \over dS} = {dN \over d\log L_{X}}{d\log L_X \over dS},
\end{equation}
where
%
%
%
%
%
\begin{equation}
\label{eq:dndlogX}
{dN \over d\log L_{X}} = {c \over H_{0}} \int_{z_{\mathrm{min}}}^{z_{\mathrm{max}}}
  {d\Phi(L_X\mathrm{(min)},z) \over d\log L_X} {d_l^2 \over (1+z)^2 (\Omega_m (1+z)^3 +
  \Omega_{\Lambda})^{1/2}} dz. 
\end{equation} 
In this last expression $d\Phi(L_X\mathrm{(min)},z)/d\log L_X$ is the hard X-ray
luminosity function, $d_l$ is the luminosity distance, and
$L_{X}\mathrm{(min)}$ is the minimum X-ray luminosity that can produce a given 1.4~GHz radio flux $S$. The radio flux $S$ is related to
the rest-frame 1.4~GHz luminosity $L_{\nu}$ by
\begin{equation}
\label{eq:kcorrect}
S={L_{\nu}(1+z)^{1+\alpha} \over 4\pi d_l^2}.
\end{equation}
The last factor in equation~\ref{eq:dnds}, $d\log L_X/dS$, cannot be expressed
analytically (especially when additional flux due to star-formation is
included), and is calculated numerically.

Thus, for a given $S$, the integral in equation~\ref{eq:dndlogX} is
calculated from $z_{\mathrm{min}}=0$ to $z_{\mathrm{max}}=5$ with a new
  value of $L_{X}\mathrm{(min)}$ for each evaluation of the
  integrand. Separate calculations are performed for radio-quiet type 1
  AGNs, radio-loud type 1 AGNs, radio-quiet type 2 AGNs, and radio-loud
  type 2 AGNs. The total 1.4~GHz $dN/dS$ is then the sum of the
  results of the four individual calculations.

\section{Results}
\label{sect:res}
\begin{figure}
\includegraphics[angle=-90,width=\textwidth]{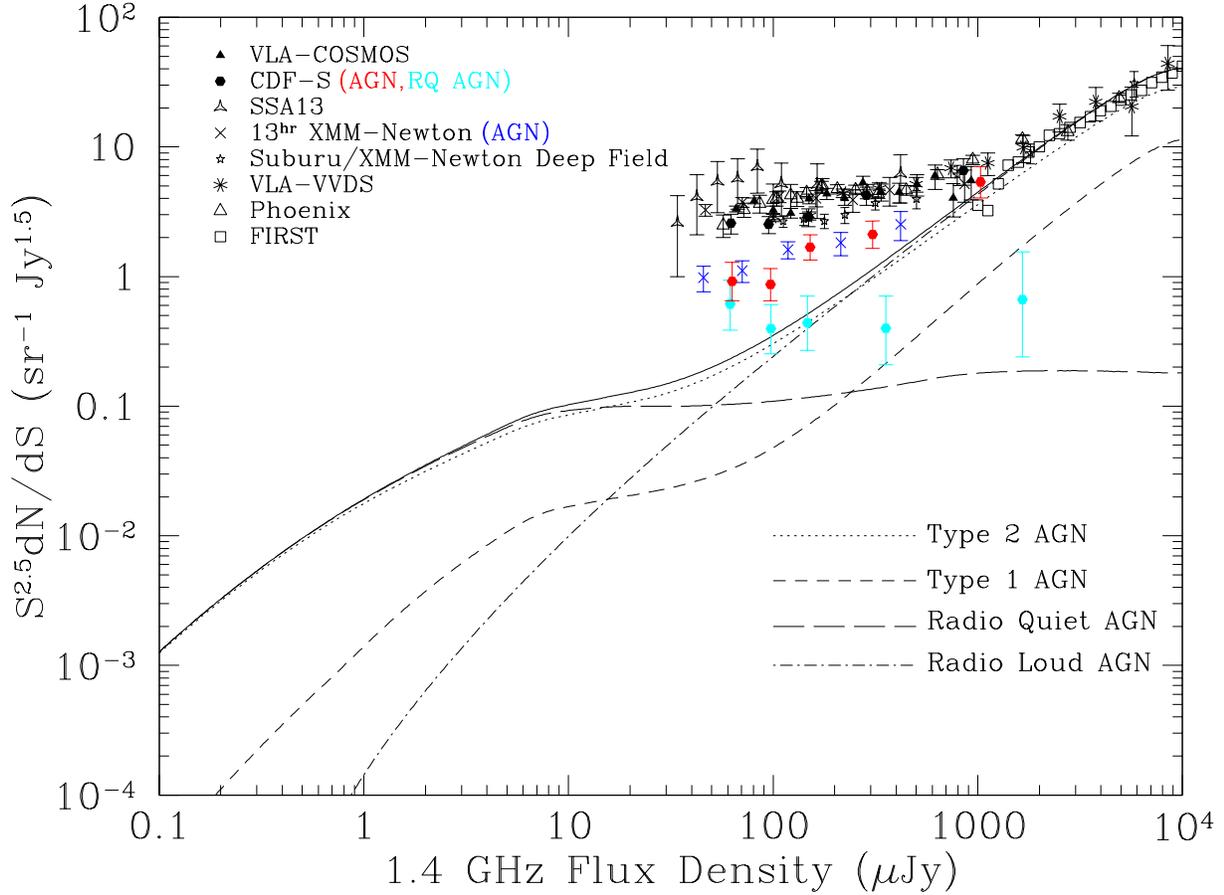}
\caption{The solid line plots the predicted Euclidean-normalized
  1.4~GHz radio counts from 0.1~\microjy\ to 10~mJy obtained from the X-ray background modeling
  described in Section~\ref{sect:calc}. The dotted and short-dashed lines
  plot the contribution to the counts from obscured and unobscured
  AGNs, respectively. These curves include AGNs of all radio powers. Type 2 AGNs dominate both the radio-loud and
  radio-quiet populations at these flux levels. The contribution from
  radio-quiet AGNs is plotted as the long-dashed line, while
  radio-loud AGNs are shown as the dot-dashed line. These curves
  include both obscured and unobscured AGNs. The points indicate
the observed counts obtained from various surveys: VLA-COSMOS \citep{bondi08},
CDF-S \citep{kell08}, SSA13 \citep{fom06}, 13hr \textit{XMM-Newton} \citep{smg04},
Suburu/\textit{XMM-Newton} Deep Field \citep{simp06}, VLA-VVDS \citep{bondi03}, Phoenix
\citep{hopk03} and FIRST \citep{white97}. The blue and red data plot the estimated
AGN contribution to the radio counts from the 13hr \textit{XMM-Newton}
\citep{sey08} and CDF-S \citep{pad09} surveys, respectively. The cyan data points are estimates of
the radio-quiet AGN contribution in the CDF-S \citep{pad09}.}
\label{fig:agntype}
\end{figure}

The solid line in Figure~\ref{fig:agntype} plots the predicted
Euclidean-normalized radio counts from AGNs at 1.4~GHz. In this figure
the contributions from obscured, type 2 AGNs and unobscured, type 1
AGNs are indicated by the dotted and short-dashed lines,
respectively. Obscured AGNs dominate the predicted radio number counts from
0.1~\microjy\ to 10~mJy by factors $\ga 3$. This result is
independent of the details of any of the modeling and is a very
general conclusion. The X-ray background requires type 2 AGNs to
outnumber type 1s at all but the highest luminosities (see
Figure~\ref{fig:batf2}), and this fact immediately translates to the
radio number counts. Therefore, AGNs detected at \microjy\ flux levels
by the EVLA or the SKA will very likely be obscured type 2 objects.

The black symbols in Figure~\ref{fig:agntype} plot the observed total 1.4 GHz number counts taken from
various published surveys over the last decade. The flattening of the
counts at fluxes below $1$~mJy is reproduced by every dataset. By construction the predicted counts pass through the majority of the
data points for $S > 1$~mJy, as radio-loud AGNs dominate this regime,
and the radio-loud fraction was determined by matching the model to
the mJy data. At lower flux densities the model passes well below the
observed data, but this is to be expected as star-forming galaxies
should dominate the \microjy\ radio counts \citep{grup03,hun05}. However, the model also
significantly underpredicts (by a factor of $\sim 5$ at 30~\microjy)
the estimated number counts due to AGNs (the red and blue data,
taken from \citet{pad09} and \citet{sey08}, respectively). Therefore, our baseline
model that is derived from an accurate fit to the X-ray background cannot
account for the observed AGN number counts in the latest deep radio surveys. 

Figure~\ref{fig:agntype} illustrates that the problem in the modeling
arises from the contribution of radio-quiet AGNs. 
Radio-loud AGNs, denoted by the dot-dashed-line in the figure, are fixed
by the mJy population and cannot help make up the deficit at $S <
100$~\microjy\ unless the radio-loud population evolves very
differently than what is currently observed. \citet{pad09} estimated
the radio-quiet AGN contribution to the 1.4 GHz number counts in the
CDF-S field, and these are plotted as the cyan data. The model
radio-quiet population underpredicts the observed estimates by
factors of $\sim 3$--$5$. 

\begin{figure}
\includegraphics[angle=-90,width=\textwidth]{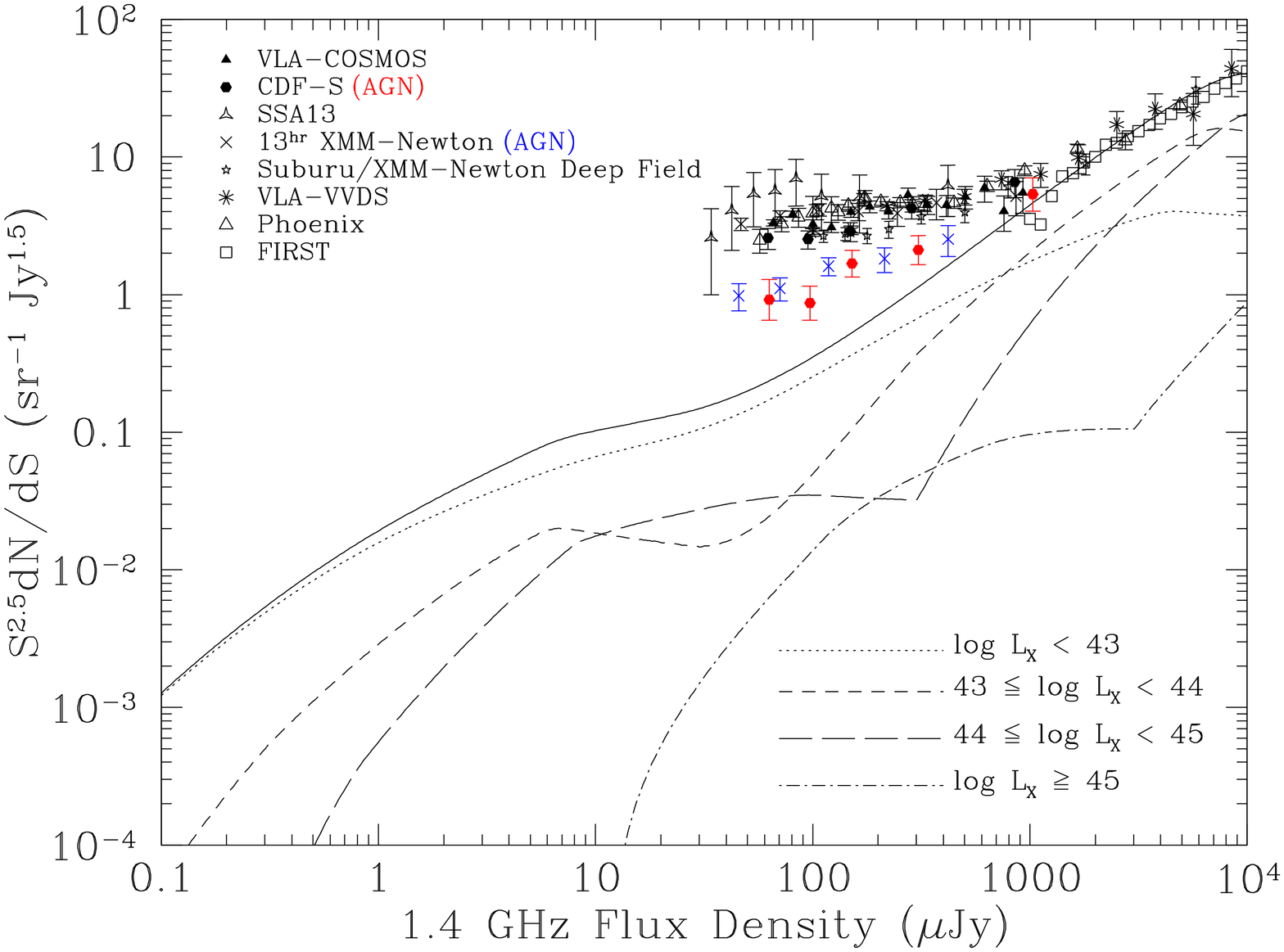}
\caption{As in Figure~\ref{fig:agntype}, but the contributions to the
  total number counts from AGNs with different absorption-corrected
  X-ray luminosities are now indicated.}
\label{fig:lumin}
\end{figure}

The breakdown of the 1.4~GHz number counts into AGNs of differing X-ray
luminosities is shown in Figure~\ref{fig:lumin}.
When $S \ga 1$~mJy, the radio counts are dominated by
moderately-luminous radio-loud AGNs with $\log L_{X} \geq 43$. The
radio-quiet population at lower flux densities is dominated by low luminosity
Seyfert galaxies. Indeed, AGNs with X-ray luminosities $\log L_{X} <
43$ comprise the largest fraction of the radio-quiet number
counts. This fact explains why obscured AGNs are the greatest
contributor to the number counts at \microjy\ fluxes.

\begin{figure}
\includegraphics[angle=-90,width=\textwidth]{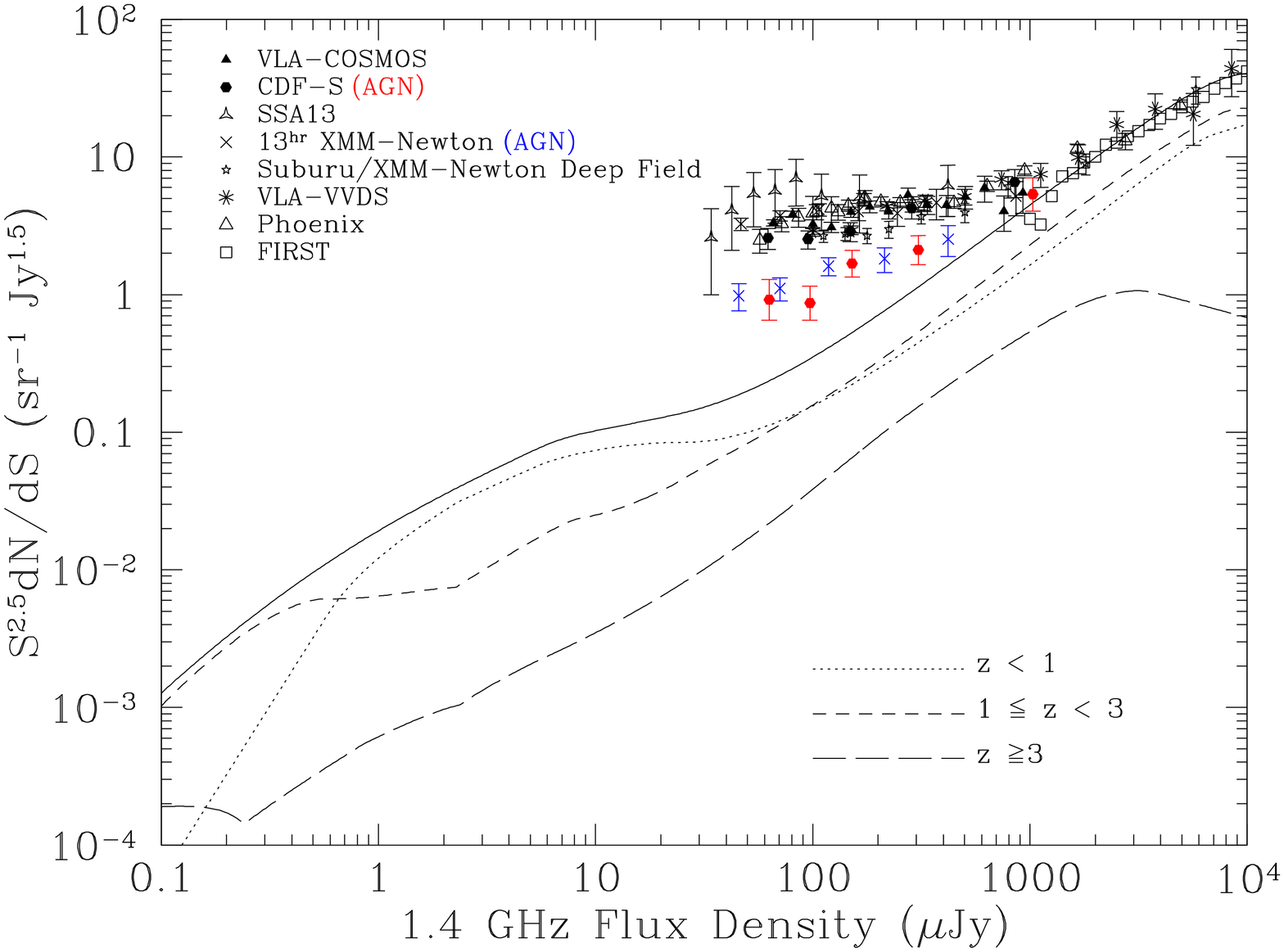}
\caption{As in Figure~\ref{fig:agntype}, but the contributions to the
  total number counts from AGNs at different redshifts are now indicated.}
\label{fig:redshift}
\end{figure}

The dependence of the AGN 1.4~GHz counts on source redshift is shown
in Figure~\ref{fig:redshift}.
Not surprisingly, the radio-loud AGNs are predominately found at $1
\leq z \leq 3$ where higher luminosity AGNs are more common. However,
once the radio-quiet population starts dominating the counts at $S \la
100$~\microjy, we find that Seyfert galaxies at $z < 1$ are the most common
contributor to the radio-quiet AGN counts for $0.5 \la S \la
100$~\microjy, but at lower fluxes the counts are dominated by
increasingly more distant AGNs. This behavior is in contrast to models
of the radio-counts where pushing to fainter and fainter flux limits
only reveals less luminous sources at roughly the same distance
\citep{cond89}. Figures~\ref{fig:lumin} and~\ref{fig:redshift} show that this
is not the case for radio surveys of AGNs. In fact, these calculations
show that \microjy\ radio observations will be a good probe of tracing
the evolution of Seyfert galaxies over a large fraction of cosmic time.

The above results have shown that the predicted 1.4~GHz radio counts
from radio-quiet AGNs are $\sim 5\times$ too low when compared against the
latest estimates from deep surveys. One way of increasing the model AGN radio counts is to add further
Compton-thick AGNs. If the additional Compton-thick AGNs had very
large columns with $\log N_{\mathrm{H}} \ga 25$ then, in principle,
the X-ray background would be insensitive to their presence. However,
the Compton-thick fraction has to be increased to 10 times the number
of Compton-thin type 2s in order to bring the radio-quiet number
counts to the level estimated by \citet{pad09}. Even if all these
Compton-thick AGNs has column densities $\log N_{\mathrm{H}} > 25$,
the peak intensity of the X-ray background spectrum would be grossly
overestimated. We conclude that a large population of hidden
Compton-thick AGNs can not account for the deficit in the predicted
1.4~GHz counts.
 
\subsection{The Role of Star Formation}
\label{sub:starformation}
A natural way to increase the radio luminosity from AGNs is to have
ongoing star-formation in the host galaxy. The fact that the AGNs
contributing to the \microjy\ radio emission have significant nuclear obscuration
is also a clue that star-formation may be important, as it may play a
significant role in obscuring Seyfert galaxies at $z \sim 1 $
\citep{fab98,wn02,tqm05,ball08}. The calibration of \citet{bell03} is used to calculate the 1.4~GHz
radio luminosity from a given SFR, and
$\alpha=-0.8$ is assumed for the radio spectrum of the star-forming
regions \citep{yrc01}. To calculate the number counts with both AGN and
star-formation emission, the star-forming 1.4~GHz flux at a given SFR and
$z$ is added to the AGN flux, and then the integral in
equation~\ref{eq:dndlogX} is computed as before. As type 2 AGNs
dominate the AGN radio counts at all fluxes (see
Fig.~\ref{fig:agntype}), emission from star-formation was only added
when calculating the total flux of obscured AGNs. Calculations
including star-formation in type~1 AGNs had no impact on the
results. Likewise, although the star-forming radio flux was added to
the radio-loud population, the AGN component easily overwhelmed the
emission from star-formation and the counts at $S \ga 1$~mJy did not
change. 

\begin{figure}
\includegraphics[angle=-90,width=\textwidth]{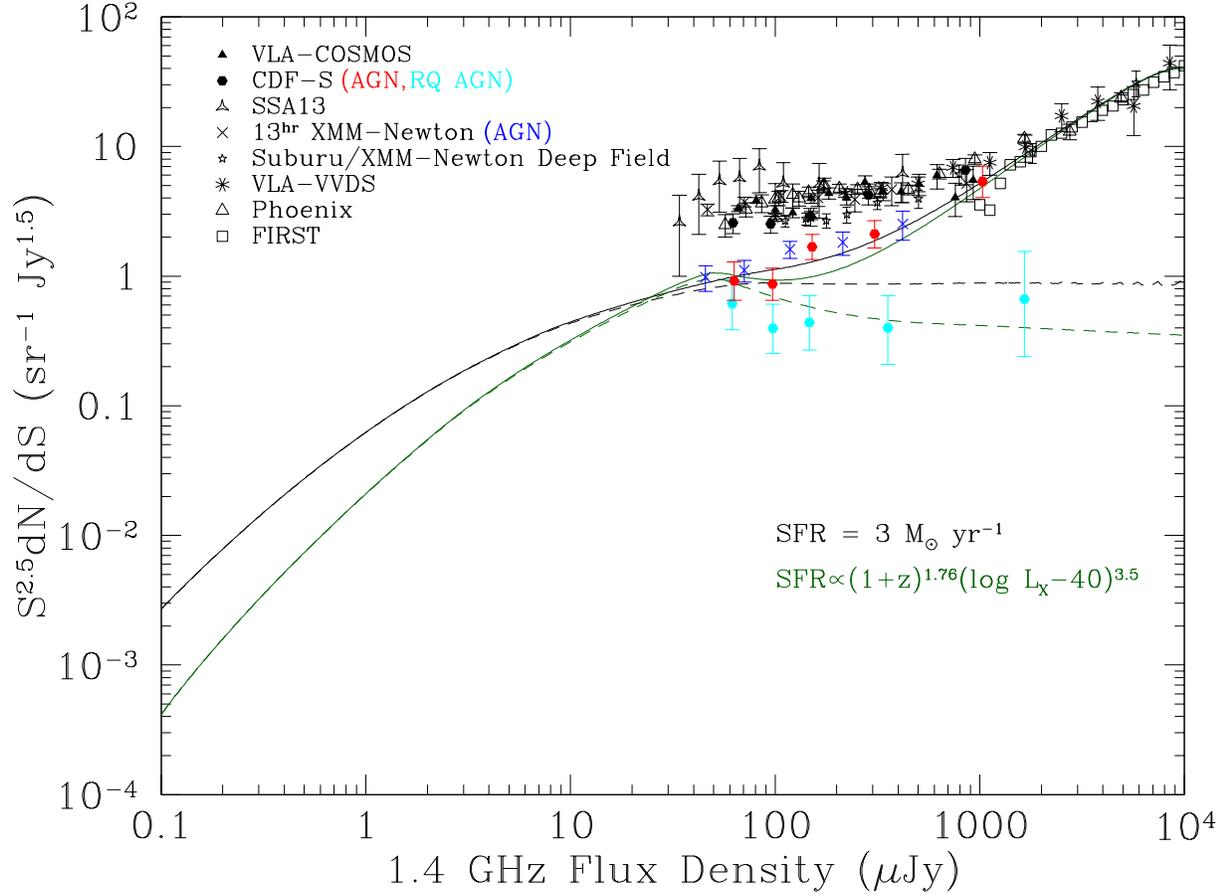}
\caption{The black solid line plots the predicted number counts from
  AGNs including radio emission from a constant SFR of $3$~\sfr. The
  black dashed line shows the contribution from radio-quiet AGNs in
  this scenario. The green lines are the analogous curves for a model
  where the SFR is a function of both $\log L_X$ and $z$ (eq.~\ref{eq:sfrlaw}). In both
  cases the radio emission was added to only the type 2 AGNs. The data
  points are the same as in Fig.~\ref{fig:agntype}.}
\label{fig:constsf}
\end{figure}

The black curves in Figure~\ref{fig:constsf} show the predicted 1.4~GHz radio counts when
a constant SFR of 3~\sfr\ was assumed for all obscured AGNs,
independent of $L_{X}$ or $z$. 
With this simple addition the model is now very close to the observed
estimates of both the total and radio-quiet AGN contribution. This
modest SFR would not be out of place in the Milky Way \citep{ss88}, and is
consistent with constraints from the cosmic infrared background
\citep{bp07}. The fact that adding such a moderate amount of star-formation in AGN
host galaxies can bring the models in line with the data is strong
evidence that such levels of star-formation is common in the host
galaxies of obscured AGNs at $z \ga 0.5$ \citep{silv09}. Moreover, the deep \microjy\
radio surveys will be able to constrain the average SFRs in obscured
AGNs and determine any dependence on luminosity and $z$.

As an illustration of what may be expected, we consider a SFR that is a
function of both redshift and the AGN luminosity. The luminosity dependence may be expected if star-forming
disks at distance $\sim 100$~pc from the central engine both obscure
and feed the black hole \citep[e.g.,][]{shi07,wki08}. In the analytic
models of \citet{ball08}, the
maximum SFR found in these disks was proportional to the X-ray
luminosity, so that less-luminous Seyfert galaxies required a lower
SFR than more luminous AGNs. The reason is that to fuel a luminous AGN
requires a significant gas supply that can also turn rapidly into
stars. Using these results as a guide, the luminosity
dependence seems to roughly follow SFR$ \propto (\log
L_{X}-40)^{3.5}$. 

To estimate the redshift dependence, we note that \citet{khi06} mention that
the SDSS type 2 AGNs identified by \citet{zak03} have a SFR$\sim 20$~\sfr\ as
judged by the [O II] line strength. The mean [O III] luminosity for
their sample is $L_{\mathrm{[O\ III]}} \approx 3\times
10^{42}$~erg~s$^{-1}$. \citet{heck05} found that for local type 1 AGNs, the average
ratio between $L_X$ and $L_{\mathrm{[O\ III]}}$ is 1.59 dex. These
authors do not correct for absorption in their sample of type 2 AGNs to find the ratio
for obscured AGNs, so we follow the AGN unified model and apply the
type 1 conversion to the \citet{zak03} sample resulting in $L_{X} \approx
10^{44}$~erg~s$^{-1}$. The majority of
the SDSS objects are at $z \sim 0.3$ \citep{zak03}. \citet{ball08} found a SFR$\sim 40$~\sfr\
in a nuclear starburst disk fueling an AGN with $L_{X} \approx
10^{44}$~erg~s$^{-1}$. If we assume that this situation is common at
$z \sim 1$ (i.e., at the peak of the cosmic star-formation history),
then the SFR in the AGN host galaxy varies as
$(1+z)^{1.76}$. Surprisingly, this is comparable to the redshift
dependence of the SFR in quasar host galaxies inferred by \citet{sh09}
using a sample selected in the far-infrared;
therefore, this redshift dependence does not seem
unreasonable. As with the AGN type 2/type 1 ratio, the redshift
evolution was halted at $z=1$.

\begin{figure}
\plotone{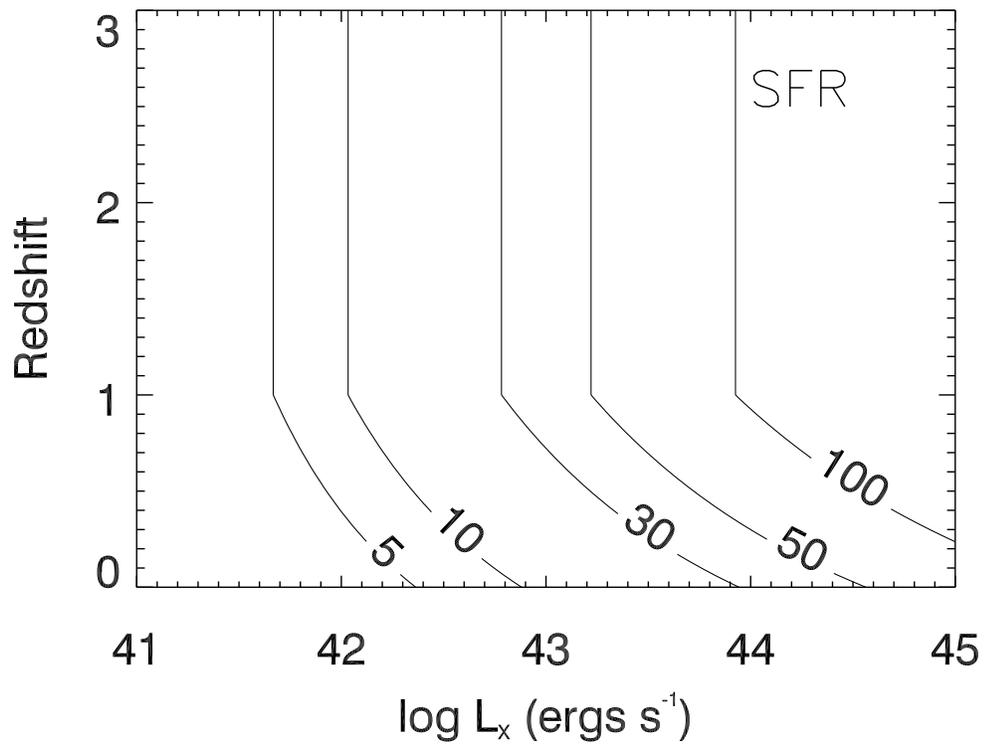}
\caption{Contours of SFR in \sfr\ as a function of $\log L_X$ and $z$
  calculated using equation~\ref{eq:sfrlaw}. This law indicates that
  SFRs $\la 30$~\sfr\ may be common in the host galaxies of the
  \microjy\ AGN population.}
\label{fig:sfrlaw}
\end{figure}

The normalization of the SFR evolution was adjusted to obtain an
adequate looking fit to the radio data. The final evolutionary law is
\begin{equation}
\mathrm{SFR} \approx 0.25(1+z)^{1.76}(\log L_{X}-40)^{3.5}\
\mathrm{M}_{\odot}\ \mathrm{yr}^{-1},
\label{eq:sfrlaw}
\end{equation}
for $z < 1$ (the SFR remains at the $z=1$ values at larger redshifts)
and is plotted in Figure~\ref{fig:sfrlaw}. The redshift and luminosity
dependences in equation~\ref{eq:sfrlaw} are steeper than those that were
recently found by \citet{silv09} by studying AGN hosts in the zCOSMOS
survey. However, the scatter in the data is about an order of
magnitude, and the predicted SFRs fall within the range observed in
the zCOSMOS data. Given the uncertainties in both the modeling
 and observations, equation~\ref{eq:sfrlaw} should be considered as a
 reasonable, observationally-motivated suggestion for how the SFR in
 AGN host galaxies may evolve.

\begin{figure}
\includegraphics[angle=-90,width=\textwidth]{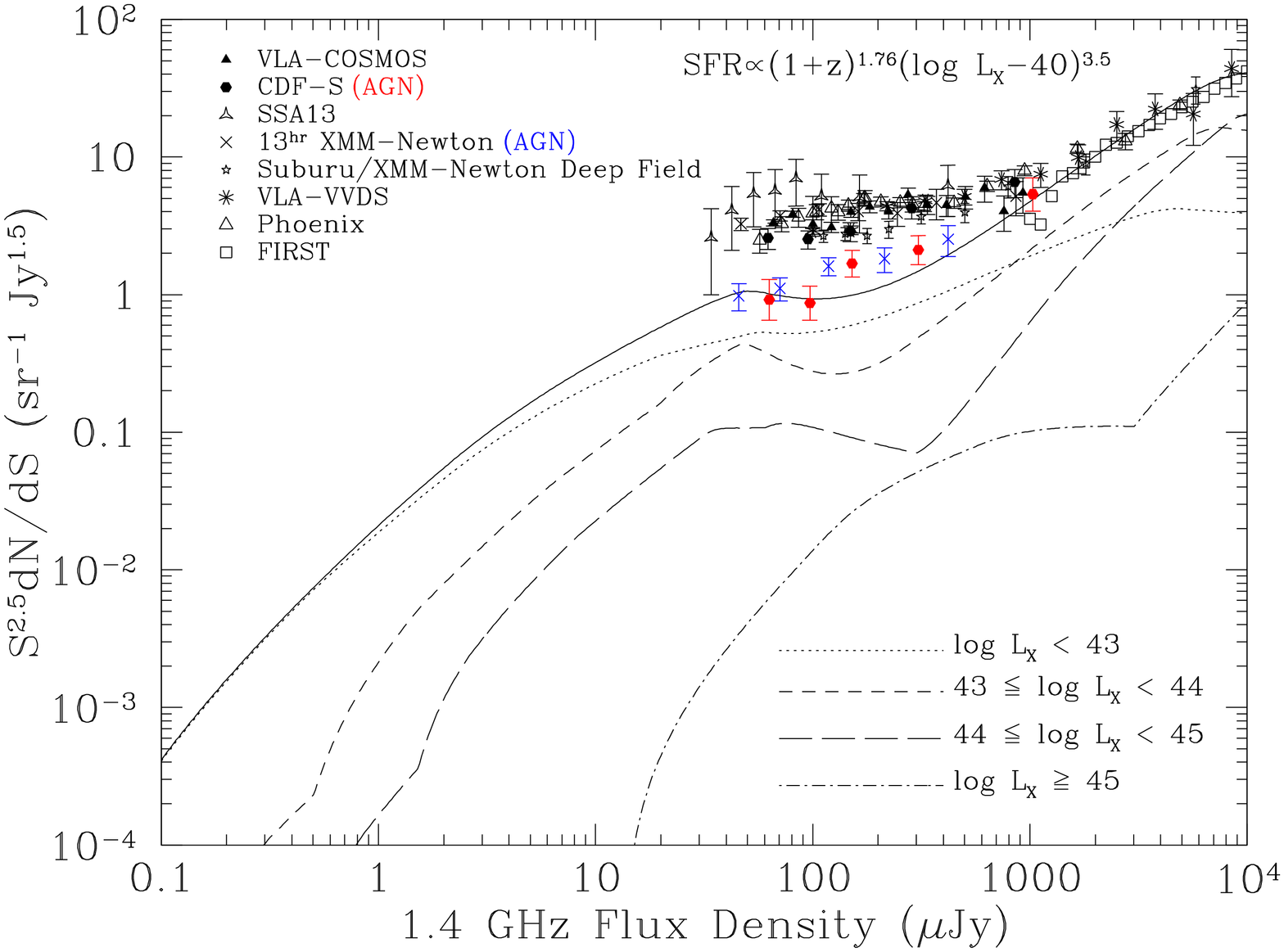}
\caption{As in Figure~\ref{fig:lumin}, but radio flux from a SFR given
  by eq.~\ref{eq:sfrlaw} was added to the type 2 AGNs.}
\label{fig:agnLsfrlaw}
\end{figure}

 The 1.4~GHz radio counts from AGNs and their host galaxies, assuming a
 SFR following eq.~\ref{eq:sfrlaw}, are shown as the green curves in Figure~\ref{fig:constsf}.
The new predicted radio-counts pass through the majority of the
observed data points. AGNs with $\log L_{X} < 43$ still provide the
majority of the counts at $S < 100$~\microjy, but higher luminosity
AGNs are now a significant contributor at $S \sim 40$~\microjy\
(Figure~\ref{fig:agnLsfrlaw}). Comparing
 Figs.~\ref{fig:agnLsfrlaw} and~\ref{fig:lumin}, we see that the
 star-formation has the largest impact on the counts only for AGNs
 with $\log L_{X} < 44$. This is because the radio-flux from the AGN
 grows faster with luminosity than the flux from the star-forming
 regions. The \microjy\ counts will then be most sensitive to
 measuring the SFR in Seyfert galaxies.
 
With the addition of the redshift-dependent star-formation law, the redshift distribution of the AGN radio-counts 
spreads to higher fluxes (see Figure~\ref{fig:zsfrlaw}).
In this model, radio-quiet AGNs at $z \leq 1$ are predicted
to be common at $S \sim 40$~\microjy, as compared to the case with no
star-formation where these AGNs peak at $S \sim 10$~\microjy\
(Fig.~\ref{fig:redshift}). The planned EVLA surveys should measure
the total radio-counts down to $\sim 1$~\microjy. Therefore, the
evolution of star-formation in AGN host galaxies should, in principle,
be measured by the deep EVLA surveys.

\subsection{Is Star Formation Required?}
\label{sub:panessa}
\begin{figure}
\includegraphics[angle=-90,width=\textwidth]{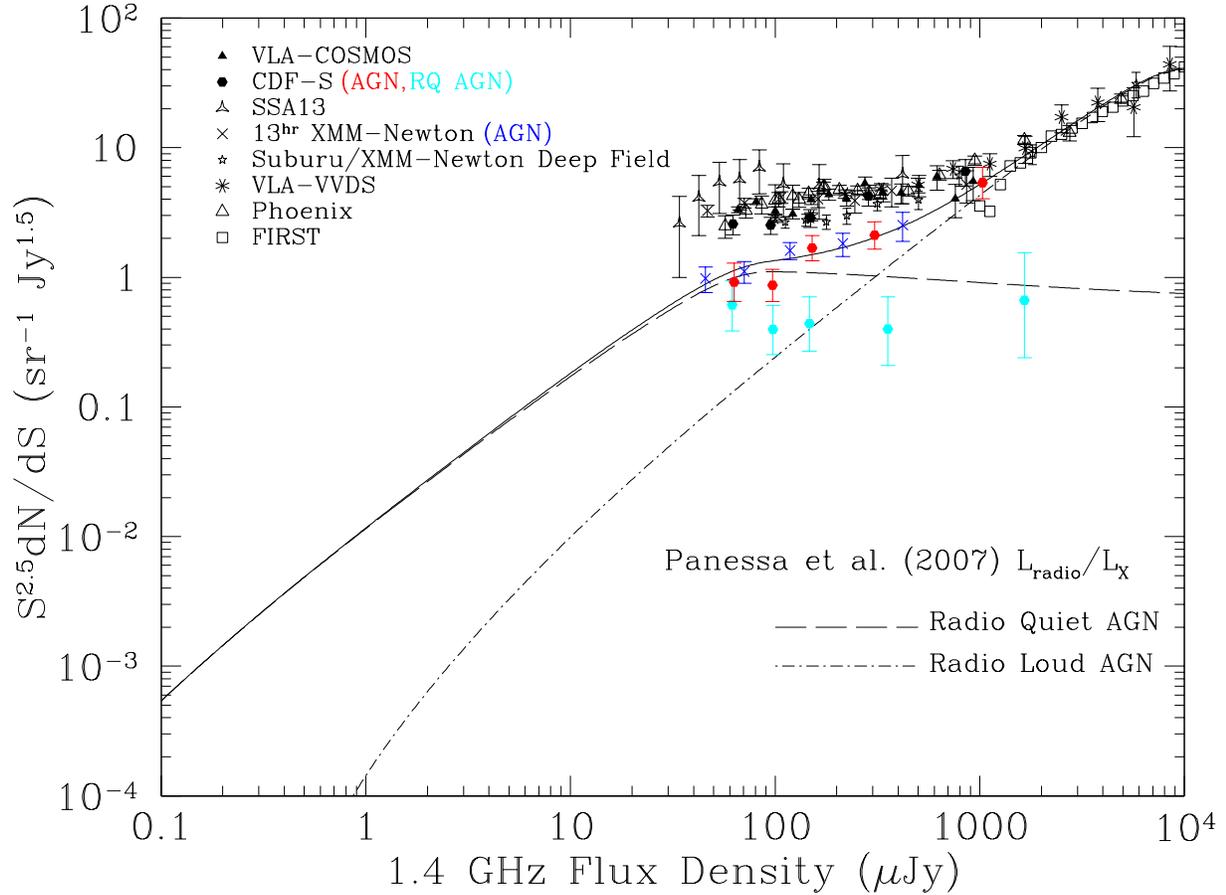}
\caption{The solid line plots the predicted number counts due to AGNs
  when assuming the \citet{pan07} relationship between $L_{X}$ and
  $\nu L_{\nu}\mathrm{(5\ GHz)}$. No additional emission from
  star-forming regions has been included. The contributions from
  radio-quiet and radio-loud AGNs are shown as indicated.}
\label{fig:panrqrl}
\end{figure}

The results presented above indicate that additional radio emission
from star-forming regions in AGN host galaxies must be included in
order for the predicted number counts to match the
observations. However, this conclusion will depend on the choice of the
X-ray to radio conversion employed in the modeling. We have argued in
Sect.~\ref{sect:res} that a luminosity-dependent conversion factor (eq.~\ref{eq:tw})
drawn from the work of \citet{tw03} was the most appropriate
choice. \citet{pad09} found that their estimates of the radio-quiet
contribution to the number counts were adequately described by a model
with a luminosity-independent conversion factor consistent with the \citet{brink00}
relation. To test the sensitivity of our conclusions we therefore
computed a new model of the 1.4~GHz radio counts using the
\citet{pan07} relationship derived from high-resolution observations
of local low-luminosity Seyfert galaxies: $\log L_{X} \approx (0.97) \log \nu L_{\nu}\mathrm{(5\
  GHz)} + 5.23$. Figure~\ref{fig:panrqrl} presents the results of this
calculation and clearly shows that this model does a good job fitting
the estimated AGN counts with no additional emission from
star-formation. The contribution from radio-quiet AGNs is slightly
overpredicted, but \citet{pad09} state that size of the error-bars are most
likely underestimated. While emission from star-formation in the AGN
host galaxies is not required to match the data while using the
\citet{pan07} relation, a small ($\sim 2$~\sfr) amount of
star-formation in the host galaxies is still consistent with the
observational constraints.

\begin{figure}
\includegraphics[angle=-90,width=\textwidth]{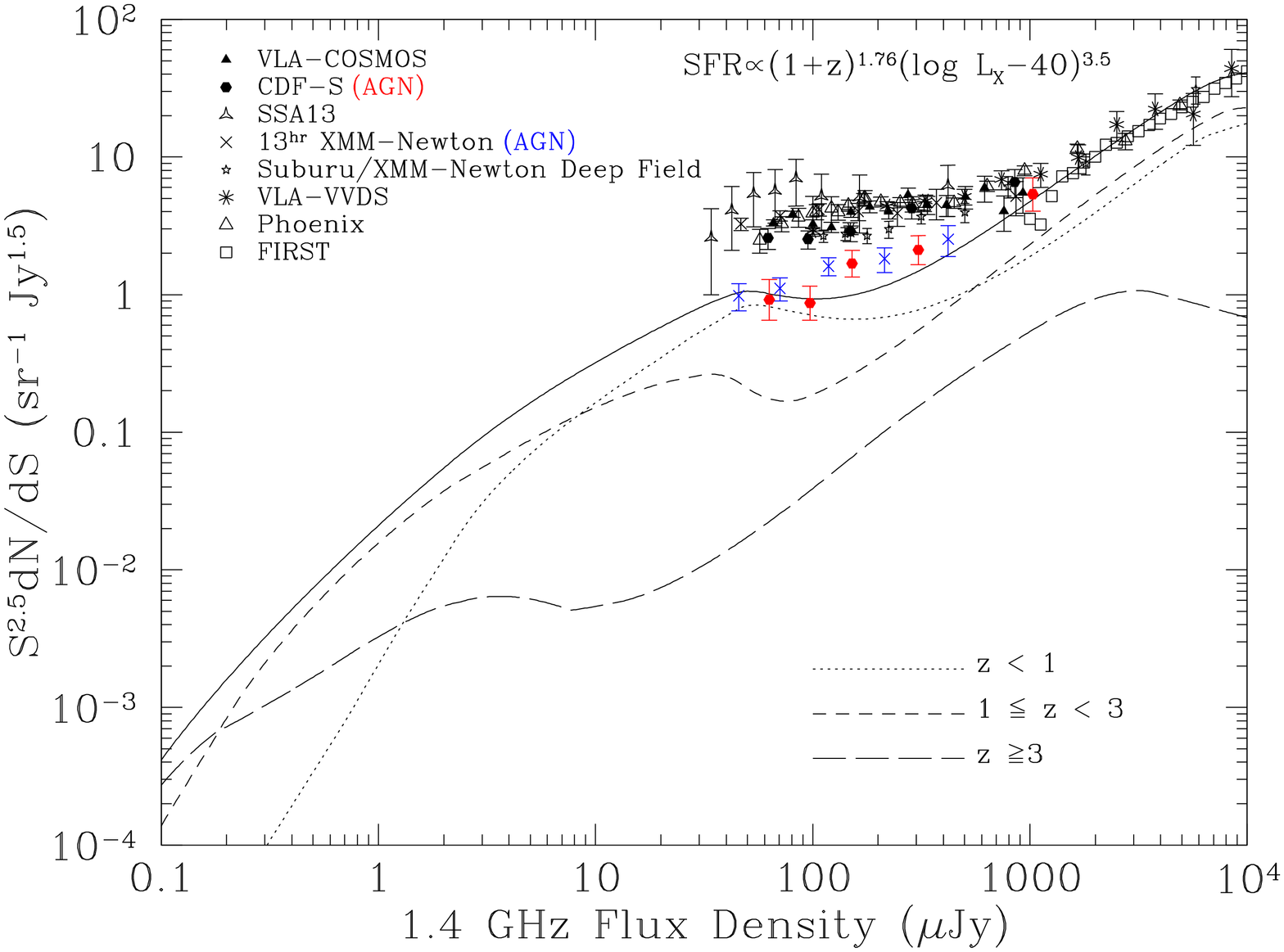}
\caption{As in Figure~\ref{fig:redshift}, but radio flux from a SFR given
  by eq.~\ref{eq:sfrlaw} was added to the type 2 AGNs.}
\label{fig:zsfrlaw}
\end{figure}

\begin{figure}
\includegraphics[angle=-90,width=\textwidth]{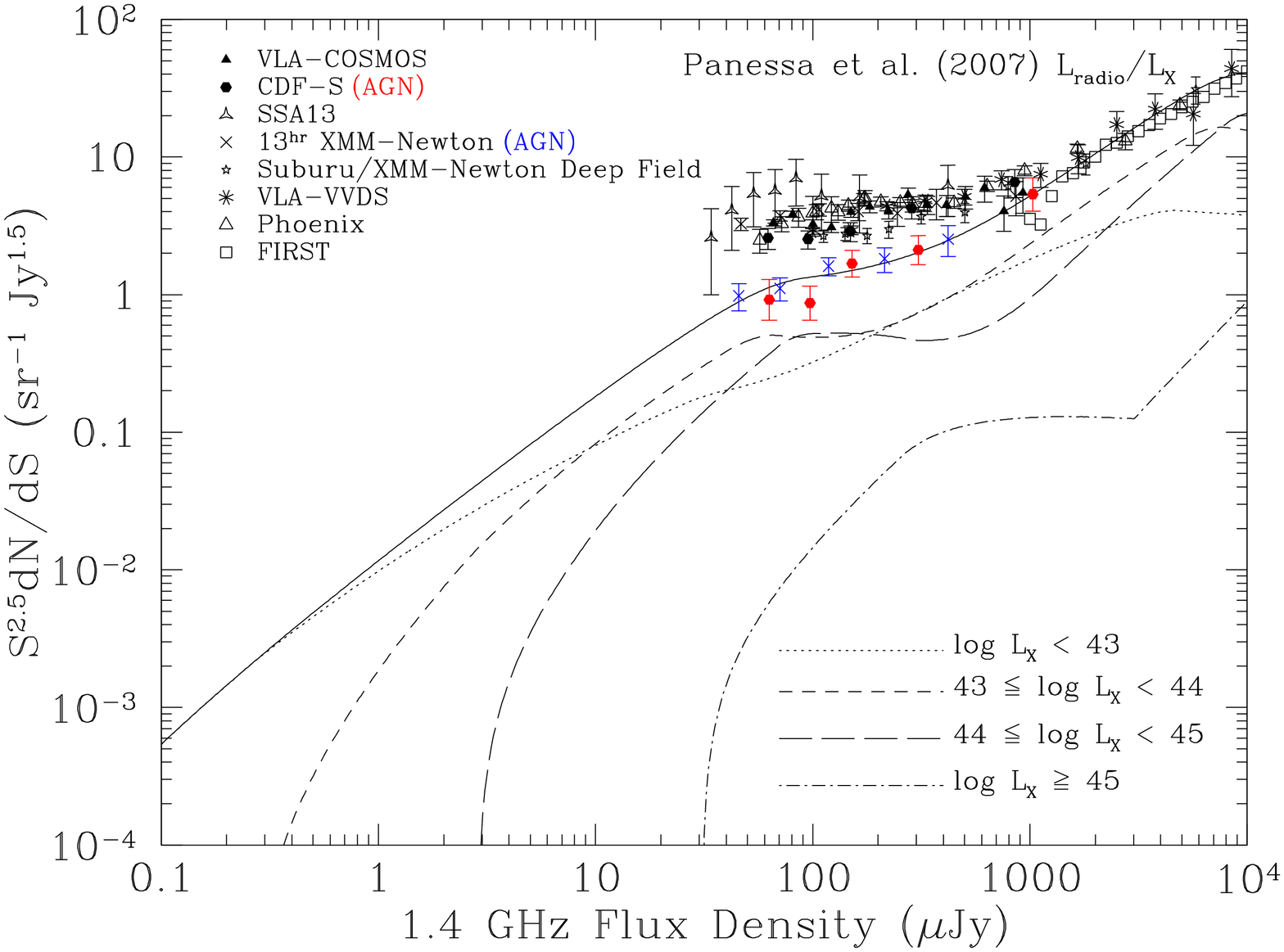}
\caption{As in Figure~\ref{fig:panrqrl}, but the contributions to the
  total number counts from AGNs with different absorption-corrected
  X-ray luminosities are now indicated.}
\label{fig:panL}
\end{figure}

\begin{figure}
\includegraphics[angle=-90,width=\textwidth]{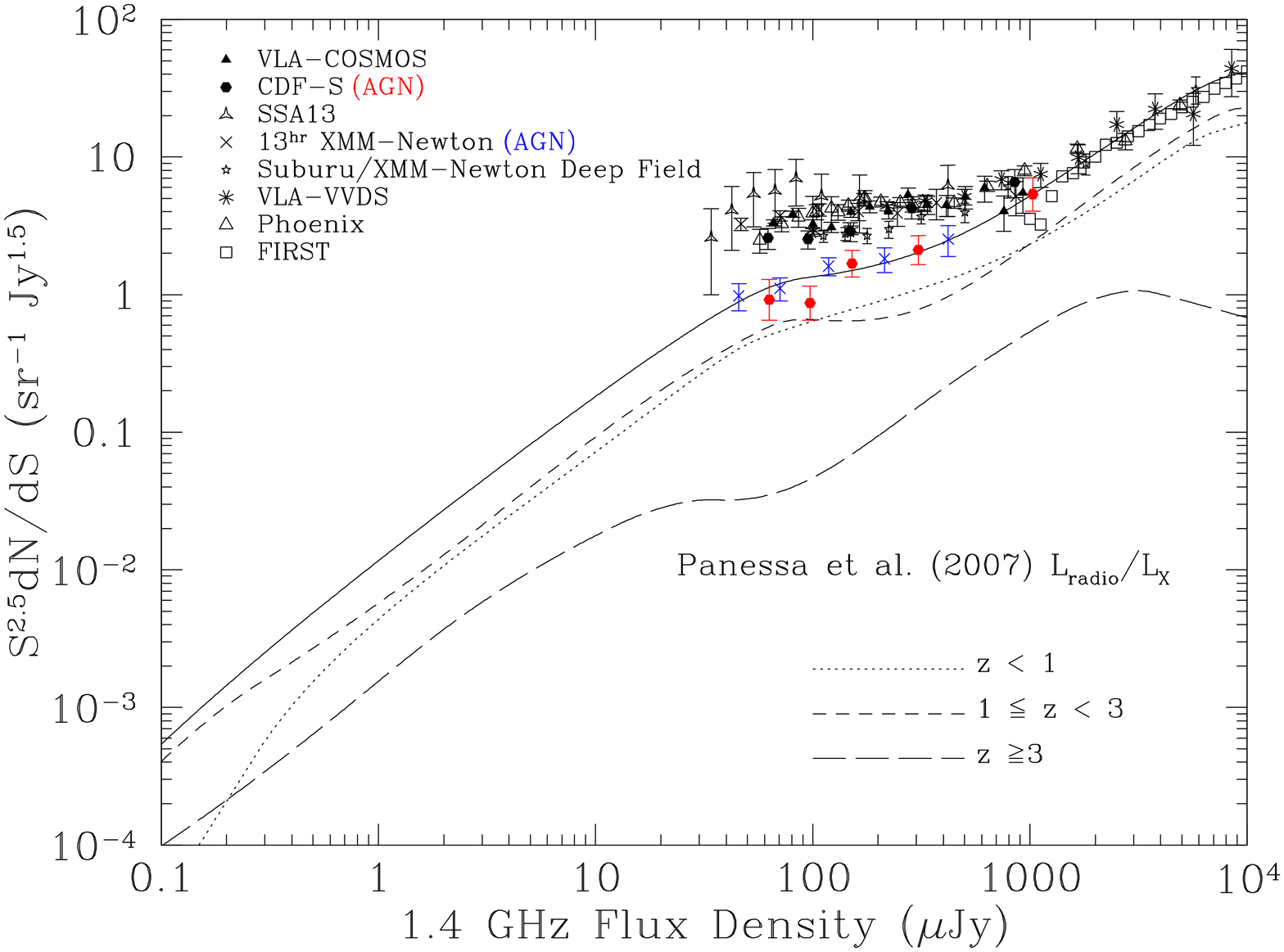}
\caption{As in Figure~\ref{fig:panrqrl}, but the contributions to the
  total number counts from AGNs at different redshifts are now
  indicated.}
\label{fig:panz}
\end{figure}

Figures~\ref{fig:panL} and~\ref{fig:panz} plot the contributions to
the counts from AGNs in different luminosity and $z$ bins in this
scenario. A comparison between these figures and
Figs.~\ref{fig:agnLsfrlaw} and~\ref{fig:zsfrlaw} show significant
differences in the properties of AGNs that dominate the \microjy\
counts. The \citet{pan07} conversion results in larger radio
luminosities for AGNs with $\log L_{X} \sim 43$--$44$ and, as these
AGNs dominate the X-ray background \citep[e.g.,][]{bem06}, they also
dominate the \microjy\ radio counts. This is in contrast with the
model that requires star-formation where AGNs with $\log L_{X} \la 43$
will be most common at these fluxes
(Fig.~\ref{fig:agnLsfrlaw}). Similarly, Fig.~\ref{fig:panz} predicts
that AGNs at $1 \leq z \leq 3$ will be equally common at \microjy\
fluxes as AGNs at $z < 1$, while Fig.~\ref{fig:zsfrlaw} shows that
AGNs at $z < 1$ will dominate if there is significant star-formation
within the host galaxies. Therefore, the luminosity and
redshift distributions of the AGNs detected in the deep radio surveys
may help indicate whether or not additional radio emission from
star-formation is necessary to explain the count
distribution. Recently, \citet{toz09} has examined the properties of
X-ray sources in the \textit{Chandra} Deep Field South that were also
detected by the deep VLA survey of that region
\citep{kell08}. \citet{toz09} found that the redshift distribution of
the radio sources with X-ray counterparts was peaked at lower
redshifts (median 0.73) than the X-ray sources without a radio match
(median 1.03). This result is qualitatively consistent with the
predictions of the AGN number counts that includes
star-formation from the host galaxy (Fig.~\ref{fig:zsfrlaw}).

\section{Discussion and Summary}
\label{sect:discuss}
The principle conclusion of this work is that, depending on the nature
of the X-ray to radio conversion in radio-quiet AGNs, the observed
radio number counts of AGNs indicate that moderate levels of
star-formation exist in the radio-quiet \microjy\ AGN population. Thus, deep
radio surveys will be an important probe of the connection between galaxy and black-hole
growth. In particular, combining \microjy\ radio data with results from
surveys at other wavelengths will probe how the star-formation
properties of AGN host galaxies evolve with $z$.

The model of the radio-counts was based on a fit to the X-ray
background spectrum and made use of the hard X-ray luminosity
function. It was found that in almost all cases the \microjy\ counts
due to AGNs are dominated by Seyfert galaxies with $\log L_{X} <
43$. The radio flux produced by more luminous AGNs was swamped by the
minority radio-loud population. The implication is that AGNs selected
by their \microjy\ radio emission will be overwhelmingly obscured
Seyfert galaxies spread over a range of redshifts. Radio selection,
therefore, will result in a way of probing the evolution of the
low-luminosity end of the hard X-ray luminosity function. This will be
especially interesting in light of recent suggestions that the
evolutionary pathways may be very different for galaxies that host
high-luminosity quasars and those that host Seyferts
\citep[e.g.,][]{bem06,has08,hick09}. The idea here is that quasars are
the signatures of the formation of massive galaxies generated by the
violent mergers of two large gas-rich galaxies
\citep[e.g.,][]{kh00,hop05}. The lower-luminosity AGNs, such as
Seyferts, do not require such large fueling rates, and may be
signatures of a less violent path of galaxy assembly. Thus, the
Seyfert galaxies found at \microjy\ radio fluxes will be a simple way
to select and measure the properties of AGNs and their host galaxies
with little contamination from the higher luminosity population.

The results described in Sect.~\ref{sect:res} show that moderate amounts of
star-formation may be ongoing in the \microjy\ AGN
population. In addition, these AGNs will be overwhelmingly obscured,
or type~2 AGNs. This is consistent with the idea that the absorbing
medium around an AGN is related to the host galaxy SFR \citep[e.g.,][]{ball08}. A
corollary of this scenario would be that the SFR and its associated
radio-flux would be correlated with X-ray column density. There are
some observational hints that this may be occurring: both \citet{georg04} and
\citet{richard07} found an increase in the number of X-ray sources with a radio
counterpart in a deep radio survey as the X-ray AGNs became
progressively more obscured (but see \citealt{toz09}). Breaking down the \microjy\ radio
counts into contributions from different X-ray obscuring columns
requires a detailed model of the circumnuclear star-forming region and
will be the subject of future work.

In summary, the increased radio continuum sensitivity of the EVLA and
the SKA will prove extremely powerful in the study of AGN
evolution. Models of the 1.4~GHz number counts derived from a fit to
the hard X-ray background show that the \microjy\ population of AGNs
will be dominated by radio-quiet, obscured Seyfert galaxies at $z \sim
1$. In addition, the radio fluxes from these AGNs may be enhanced by
galactic star-formation. Indeed, current estimates of the AGN
contribution to the 1.4~GHz radio counts seems to require moderate levels of
star-formation in order for the models to match the observed data points. Multiwavelength
followup observations of AGNs selected by their \microjy\ radio
emission will be able to trace the evolution of Seyfert galaxies from
$z \sim 3$ to the present day.

\acknowledgments
The author thanks V. Smol\v{c}i\'{c}, P. Padovani, N. Seymour and
F. Bauer for useful conversations and comments. The anonymous referee
is acknowledged for a detailed and constructive report.

{}

\newpage

\newpage

\newpage

\newpage

\newpage

\newpage

\newpage

\newpage

\newpage

\newpage

\newpage

\end{document}